\documentclass[10pt,letterpaper]{article}
\usepackage[top=0.85in,left=2.75in,footskip=0.75in]{geometry}
\usepackage{amsmath,amssymb}
\usepackage{changepage}
\usepackage{textcomp,marvosym}
\usepackage{cite}
\usepackage{nameref,hyperref}
\usepackage[nopatch=eqnum]{microtype}
\usepackage{float}
\usepackage[section]{placeins}  
\DisableLigatures[f]{encoding = *, family = * }
\usepackage[table]{xcolor}
\usepackage{array}
\usepackage{lastpage,fancyhdr,graphicx,epstopdf}
\fancyheadoffset[L]{2.25in}
\fancyfootoffset[L]{2.25in}
\raggedright
\usepackage[aboveskip=1pt,labelfont=bf,labelsep=period,justification=raggedright,singlelinecheck=off]{caption}

\makeatletter
\renewcommand{\@biblabel}[1]{\quad#1.}
\makeatother

\begin{document}
\vspace*{0.2in}

\begin{flushleft}
{\Large
\textbf\newline{Investigating Audience Preferences Within the Hybrid Competitive-Comedic Format of Taskmaster UK}
}
\newline
\\
David H. Silver\textsuperscript{1*}
\\
\bigskip
\textbf{1} Remiza AI\\
\bigskip
* Corresponding author: david@remiza.ai
\end{flushleft}

\section*{Abstract}
\textbf{Background:} Hybrid entertainment formats combining competitive and comedic elements present opportunities to investigate factors driving audience engagement. I analyzed \textit{Taskmaster UK} (2015--2023), a BAFTA-winning comedy panel show where comedians compete in creative tasks judged by a host, to quantify relationships between scoring mechanics, performer characteristics, and viewer ratings.

\textbf{Methods:} I analyzed 154 episodes encompassing 917 tasks performed by 90 contestants, with audience reception measured through 32{,}607 IMDb votes. To capture scoring dynamics while avoiding intercorrelated metrics, I employed a low-dimensional representation using mean ($\mu$) and variance ($\sigma^2$) of score distributions. Additional methods included mixture modeling for rating distributions (tri-peak model: $a_1\cdot\delta(1) + a_{10}\cdot\delta(10) + a_{\text{gaussian}}\cdot\mathcal{N}(\mu,\sigma)$), hierarchical clustering for performance patterns, and Random Forest regression. All $p$-values include False Discovery Rate correction.

\textbf{Results:} Low-dimensional scoring representation showed no significant associations with IMDb ratings ($\mu$: $r = -0.012$, $p = 0.890$; $\sigma^2$: $r = -0.118$, $p = 0.179$; combined $R^2 = 0.017$, $p = 0.698$). Contestant age emerged as the strongest predictor (39.5\% $\pm$ 2.1\% feature importance). Sentiment analysis identified increased awkwardness over time ($\beta = 0.0122$, adjusted $p = 0.0027$). Clustering revealed five performance archetypes appearing consistently across series. Geometric analysis showed 38.9\% (98/252) of mathematically possible scoring distributions occur in practice.

\textbf{Conclusions:} Competitive elements provide framework while audience engagement correlates with performer characteristics and emotional content. The low-dimensional scoring analysis eliminates methodological concerns about metric intercorrelation. These findings position \textit{Taskmaster UK} as a quantifiable example where secondary mechanics enable but do not determine primary value.


\section*{Introduction}

Hybrid entertainment formats are an increasingly prominent part of global media ecosystems. These formats combine elements of structured gameplay with unscripted humor, narrative tension, or performative expression. As such, they blur conventional genre lines, merging the competitive arc of game shows with the affective and improvisational qualities of comedy or reality television. This fusion reflects broader trends in post-network television, where satire, personality, and affect play central roles in shaping engagement and meaning~\cite{Gray2009}.

Understanding audience engagement with hybrid formats requires theoretical frameworks that account for both motivational drivers of media consumption and viewer-content relationships. Uses and Gratifications Theory provides a foundational approach, positing that audiences actively select media to satisfy specific psychological and social needs~\cite{Katz1973,Rubin2009}. In competition-based entertainment, viewers may seek gratifications ranging from social comparison and vicarious achievement to escapism and humor~\cite{Nabi2003}. Analysis of reality talent shows demonstrates how producers and audiences co-create engagement through relational dynamics that extend beyond simple consumption, with viewers moving between passive viewing and active participation~\cite{Hill2017}. Research on parasocial relationships reveals how audiences form meaningful connections with television personalities, creating sustained engagement across multiple seasons~\cite{Enli2012}.

Hybrid entertainment formats represent a significant evolution in post-network television, where traditional genre boundaries dissolve in favor of cross-genre combinations blending competitive, comedic, and reality elements~\cite{Vassallo2016,Gil2012}. Unlike traditional game shows driven primarily by competition, hybrid formats must sustain audience interest through multiple, potentially competing appeals. This requires deliberate choices that support both competitive tension and comedic improvisation~\cite{OBITEL2018}. The global proliferation of such formats has created opportunities for comparative analysis across cultural contexts~\cite{Larkey2016}, where cultural proximity theory suggests audiences prefer content reflecting their own values and humor styles~\cite{Straubhaar2007,Trepte2008}. However, successful format adaptations demonstrate that certain hybrid structures can transcend cultural boundaries when appropriately localized. Cross-cultural studies reveal significant variations in how audiences evaluate entertainment content, with differences in humor appreciation, competitive attitudes, and parasocial relationship formation~\cite{YooEtAl2014,GarciaBéjar2021}. These findings indicate that while competitive mechanics may be universal, reception processes remain culturally contingent.

Despite their growing prevalence and popularity, the mechanisms by which hybrid formats sustain viewer engagement have received limited systematic study.

\textit{Taskmaster UK}, a BAFTA-winning panel-format television show that aired from 2015 to 2023, offers an unusually well-suited dataset for addressing this gap. Each episode features five celebrity contestants (usually comedians) attempting a series of whimsical tasks. These are scored on a 0–5 scale by a host-judge and tallied to produce a running competition across a season. Although the program follows a rigid format, its tone and emphasis remain comedic. In this sense, the competitive framework exists primarily to support the humor and personality-driven interactions that constitute the show’s appeal.

This combination of consistency and creativity yields a highly analyzable structure. Across 154 episodes and 18 series, the show contains 917 discrete scoring events. Alongside this, IMDb user ratings provide a parallel signal of audience engagement: over 32,600 viewer votes are available at the episode level. These vary meaningfully across time and series, creating an opportunity to test which competitive mechanics, emotional, or performer-level factors influence reception.

\begin{figure}[!h]
\centering
\begin{minipage}{0.48\linewidth}
  \centering
  \includegraphics[width=\linewidth]{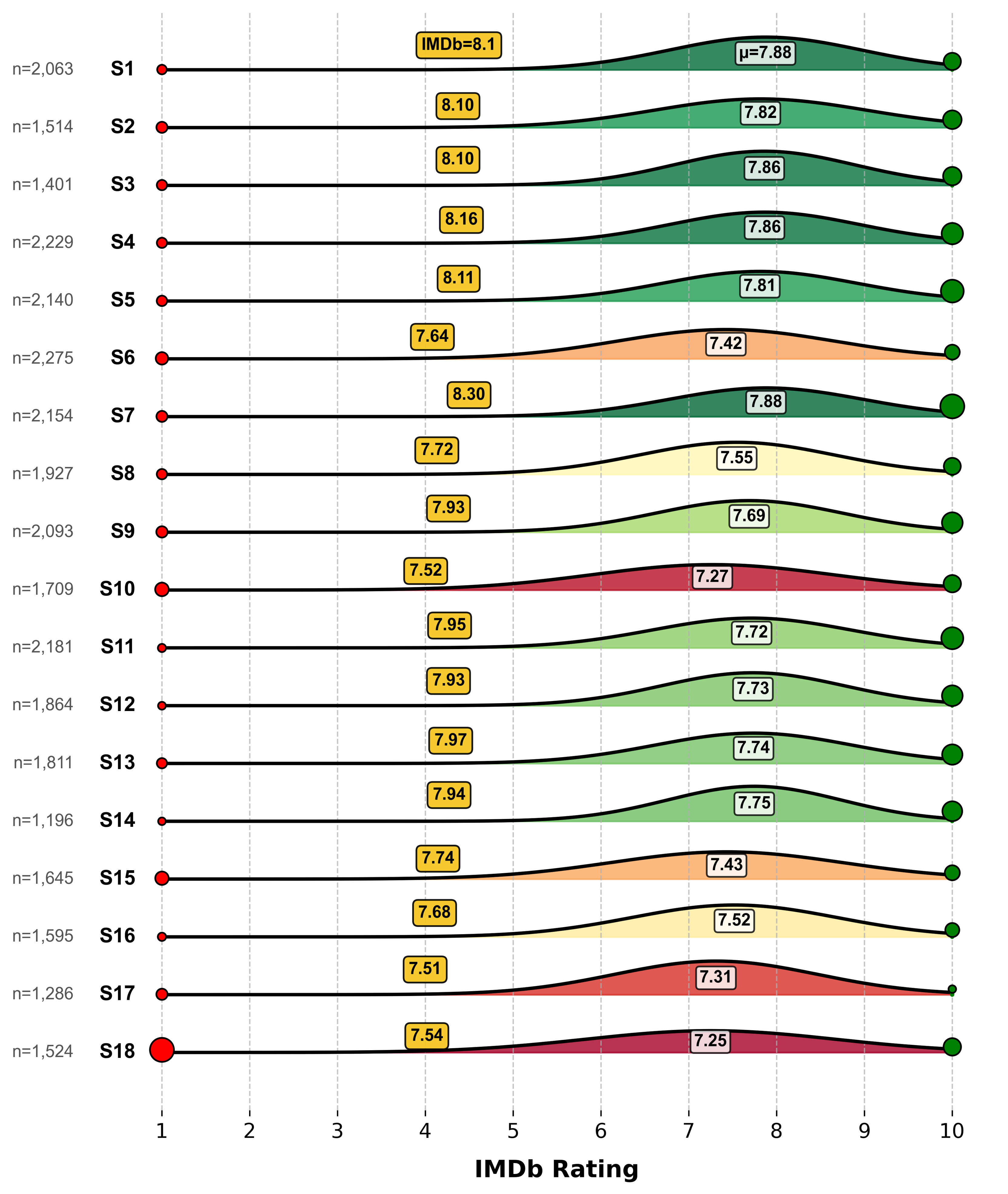}
  \vspace{1mm}
  \textbf{A}
\end{minipage}
\hfill
\begin{minipage}{0.48\linewidth}
  \centering
  \includegraphics[width=\linewidth]{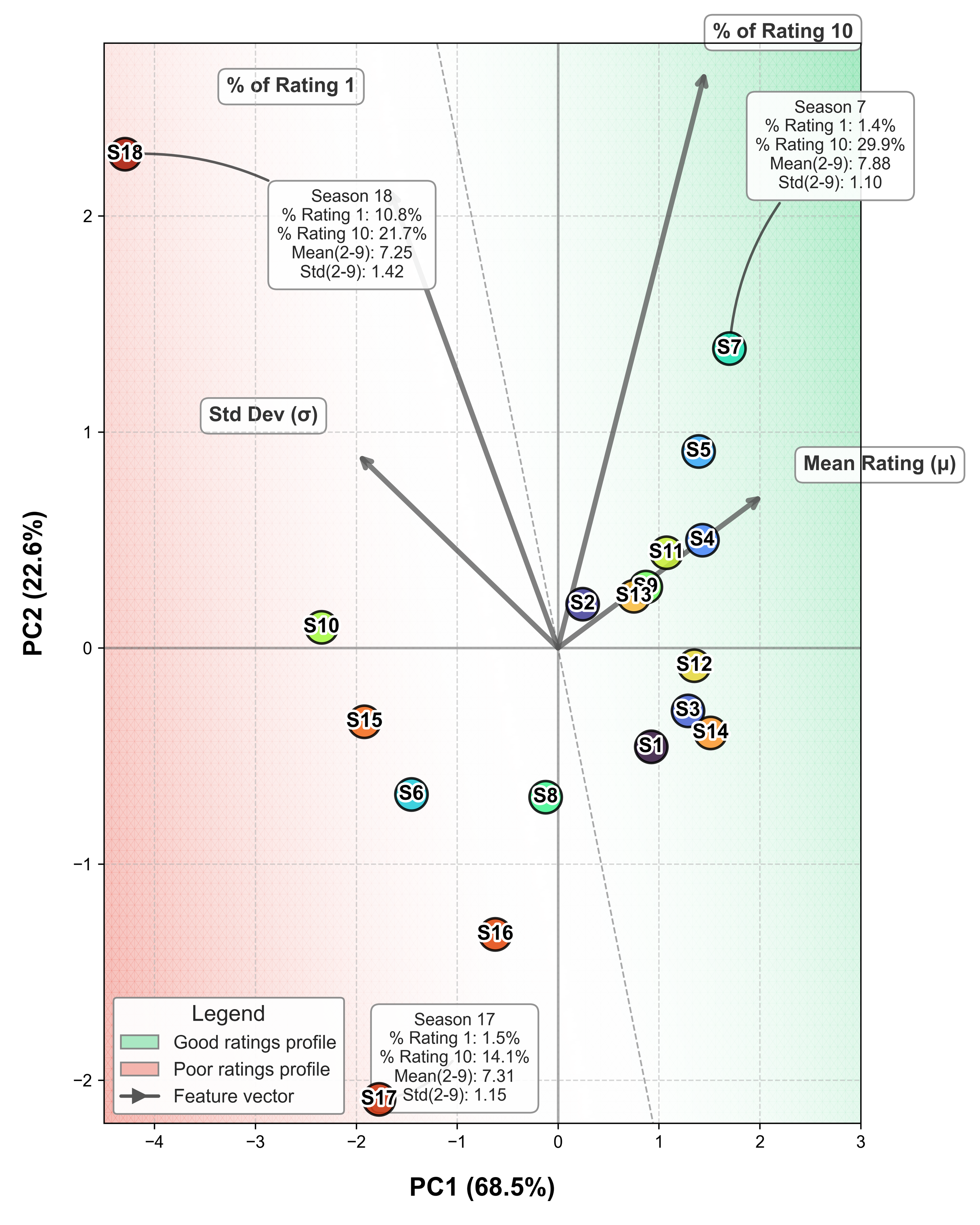}
  \vspace{1mm}
  \textbf{B}
\end{minipage}
\caption{{\bf Viewer Rating Structure Across Series.}
\textbf{Panel A:} IMDb vote distributions decomposed into a tri-modal mixture model for each series. Red and green spikes indicate the proportion of 1-star and 10-star ratings respectively, while the Gaussian curve captures central ratings (2–9). Series are labeled with official IMDb scores. 
\textbf{Panel B:} Principal Component Analysis of series-level statistics: mean and standard deviation of Gaussian ratings, percent 1s, and percent 10s, reveals a quality–polarization space that separates high-rated, controversial, and consensus series.}
\label{fig:series_ratings}
\end{figure}

To understand these dynamics, I begin with an empirical observation about the ratings themselves. Unlike traditional unimodal distributions expected from neutral audience consensus, \textit{Taskmaster} episodes tend to exhibit a distinctive tri-peak profile. A subset of viewers consistently rates episodes a perfect 10, while another subset assigns the lowest possible score (1). The remainder cluster around a broad, approximately normal distribution of mid-range ratings. This trimodal shape, modeled as a weighted combination of two delta functions and a Gaussian ($a_1 \cdot \delta(1) + a_{10} \cdot \delta(10) + a_{\text{gaussian}} \cdot \mathcal{N}(\mu, \sigma)$), is consistent with statistical mixture modeling approaches used to capture latent subpopulations within audience response data~\cite{McLachlan2000}. Fig~\ref{fig:series_ratings} illustrates these patterns and reveals how series differ along orthogonal axes of quality and polarization.

These descriptive findings motivate a broader set of theoretical questions. In other competitive formats, engagement is often driven by suspense and outcome uncertainty. Sports broadcasts, for example, attract 15--20\% higher viewership when contests are closely scored~\cite{Rottenberg1956,Fort2003}. Reality television shows receive increased ratings when eliminations are unpredictable~\cite{Hall2009}, and game shows with volatile scoring dynamics are better at retaining viewer attention~\cite{Vorderer2004}. These precedents suggest that score-related variables such as mean, variance, or temporal spread, could influence audience reactions.

Yet hybrid formats may operate under different principles. Contemporary research on digital media consumption demonstrates that audiences engage with entertainment content through complex motivational frameworks that extend beyond traditional suspense or outcome uncertainty~\cite{GarciaBéjar2021}. In \textit{Taskmaster}, the competition provides structure, but the central appeal lies in character, humor, and improvisational responses. From this perspective, viewers may be primarily seeking entertainment, social interaction, and parasocial relationships with performers rather than competitive excitement. If so, then scoring dynamics would have limited predictive value for viewer ratings, and audience engagement would depend more on performer characteristics and emotional content.

Complicating matters further, prior entertainment research often uses multiple overlapping metrics (e.g., volatility, variance, spread), which can suffer from multicollinearity and inflate false positives~\cite{Gelman2007}. To avoid this, I adopt a low-dimensional approach to scoring geometry: each task’s score distribution is summarized by its mean ($\mu$) and variance ($\sigma^2$), which jointly characterize the bulk of variation across episodes. This minimal representation uniquely distinguishes over 90\% of actual scoring patterns used in the show.

These methodological considerations lead to three pre-registered hypotheses:

\begin{enumerate}
  \item Scoring dynamics, as quantified by $(\mu, \sigma^2)$, will not significantly predict episode-level IMDb ratings. The scoring system acts as a framework rather than a driver of enjoyment.
  \item Contestant characteristics—such as age, professional experience, and comedic background—will correlate with higher ratings. These reflect performance polish and on-screen charisma.
  \item Emotional tone, derived via sentiment analysis of episode transcripts, will show consistent temporal trends. These may reflect shifts in comedic style or production strategy over the show's lifespan.
\end{enumerate}

The rest of this paper tests these hypotheses using metadata, full-episode sentiment annotations, and predictive models. By analyzing how competitive mechanics, performer traits, and emotional tone interact to shape viewer response, I offer a data-driven view of what makes hybrid entertainment compelling.

\section*{Materials and methods}

\subsection*{Dataset Overview}

This study analyzes all 18 series of \textit{Taskmaster UK} (2015–2023), comprising 154 episodes and 917 distinct tasks. Each episode features five contestants, primarily comedians, who complete various tasks scored on a 0–5 point scale. In total, the dataset includes over 4,500 individual task-level score entries. These were joined with viewer ratings, sentiment annotations, contestant demographics, and task metadata to form a unified analytical corpus.

\paragraph{IMDb Ratings and Vote Distributions}
Episode-level IMDb ratings were manually collected from IMDb’s public website and cross-referenced with official episode listings \cite{imdb}. These ratings span from 6.68 to 8.86, with a mean of 7.95 across all episodes. In addition to the official weighted averages, I extracted full vote histograms for each episode, recording the percentage of ratings at each integer level from 1 to 10. These distributions were modeled using a tri-modal mixture: delta functions at 1 and 10, and a Gaussian over the intermediate range (2–9). This model reduced mean absolute error by 48.1\% relative to a single Gaussian fit, offering a more accurate representation of polarized audience responses~\cite{McLachlan2000}.

\paragraph{Task and Episode Metadata}
Task and episode metadata were sourced from taskmaster.info, a community-maintained database that provides annotations for all 154 episodes and 917 unique tasks \cite{taskmasterinfo}. Each task was labeled along multiple dimensions, including activity type (creative, physical, mental, social), judgment type (objective or subjective), assignment format (solo, team, split), modality (filmed, live, prize, homework), and physical location. Among these, 43.5\% of tasks were coded as creative, 48.1\% as physical, and 55.9\% employed objective scoring. Solo tasks were the most common, comprising 87.9\% of the total.

\begin{figure}[!h]
\centering
\includegraphics[width=\linewidth]{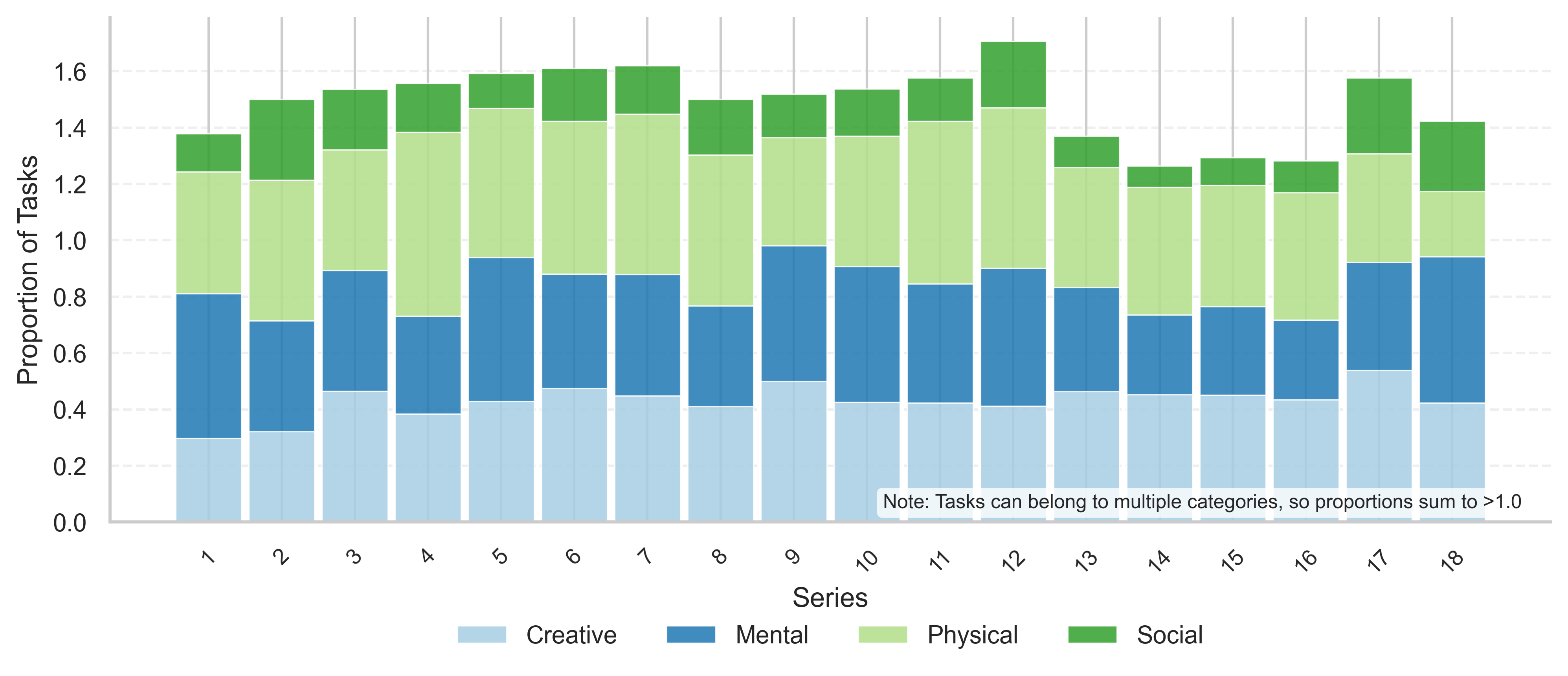}
\caption{{\bf Task Type Proportions by Series.}
The frequency of each task type remains relatively unchanged over time. No consistent trend is observed across the 18 series, underscoring the show's adherence to a stable format.}
\label{fig:task_type_stability}
\end{figure}

To assess whether this distribution evolved over time, I analyzed the frequency of each task type across the show's 18 series. As shown in Fig~\ref{fig:task_type_stability}, the relative proportions of Creative and Physical, as well as Objective and Subjective tasks, remained remarkably stable. This consistency suggests a deliberate design philosophy that balances familiarity with creative variability.

\FloatBarrier

\begin{figure}[!h]
\centering
\includegraphics[width=\linewidth]{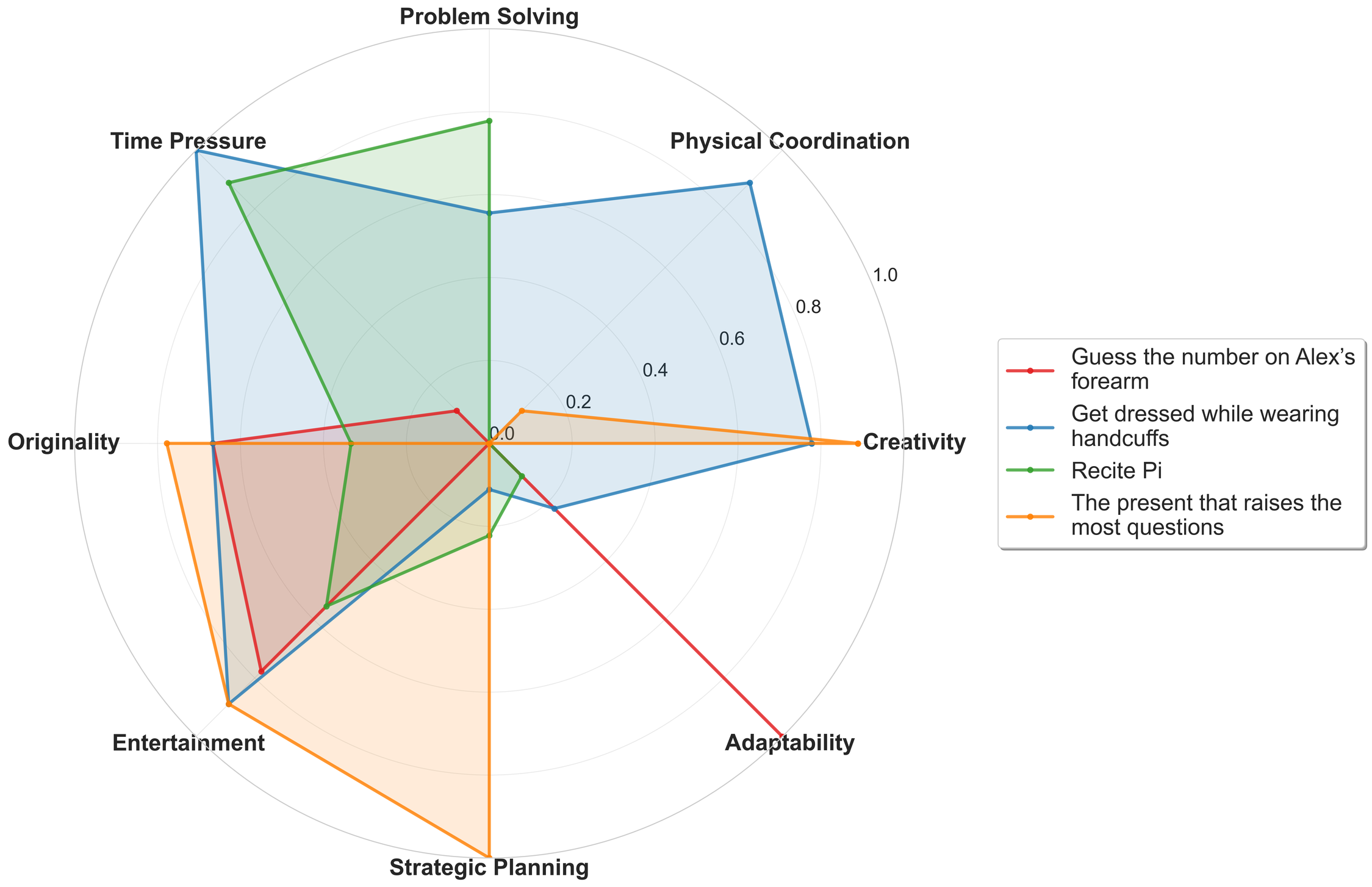}
\caption{{\bf Task Skill Intensity Profiles.}
Spider plots of four polarized Taskmaster tasks based on GPT-4o ratings along eight skill dimensions. Each task shows a distinct cognitive or physical emphasis, illustrating the diversity of design space. These plots are used for illustrative purposes only and were not included in statistical models.}
\label{fig:task_skills}
\end{figure}

To complement these categorical descriptors, I applied GPT-4o to extract continuous-valued skill profiles for each task. The model rated tasks along eight dimensions: creativity, physical coordination, problem solving, time pressure, originality, entertainment value, strategic planning, and adaptability. These machine-generated annotations allowed us to identify a subset of highly polarized tasks with distinct cognitive and physical demands (Fig~\ref{fig:task_skills}). The selection of eight dimensions reflects a balance between expressive richness and interpretability, consistent with best practices in statistical learning~\cite{Hastie2009,James2013}. Due to known limitations in large language model generalization, accuracy, and reproducibility, these outputs were not used for statistical inference or hypothesis testing. Instead, they were used to support exploratory visualization and qualitative example selection.

\FloatBarrier

The use of LLMs in this context demonstrates their value for large-scale descriptive tagging. It also underscores the importance of caution when interpreting their outputs in formal quantitative analyses.

\paragraph{Contestant Demographics}
Demographic information for all 90 contestants was compiled manually from publicly available sources, including Wikipedia, press interviews, and official biographies. For each individual, I recorded age at time of airing, gender, country of birth, and occupation category (e.g., comedian, actor, presenter). These attributes were used to compute derived features such as the average age per episode, average years of entertainment experience, and the proportion of contestants with comedy-specific backgrounds.

\begin{figure}[!h]
\centering
\includegraphics[width=\linewidth]{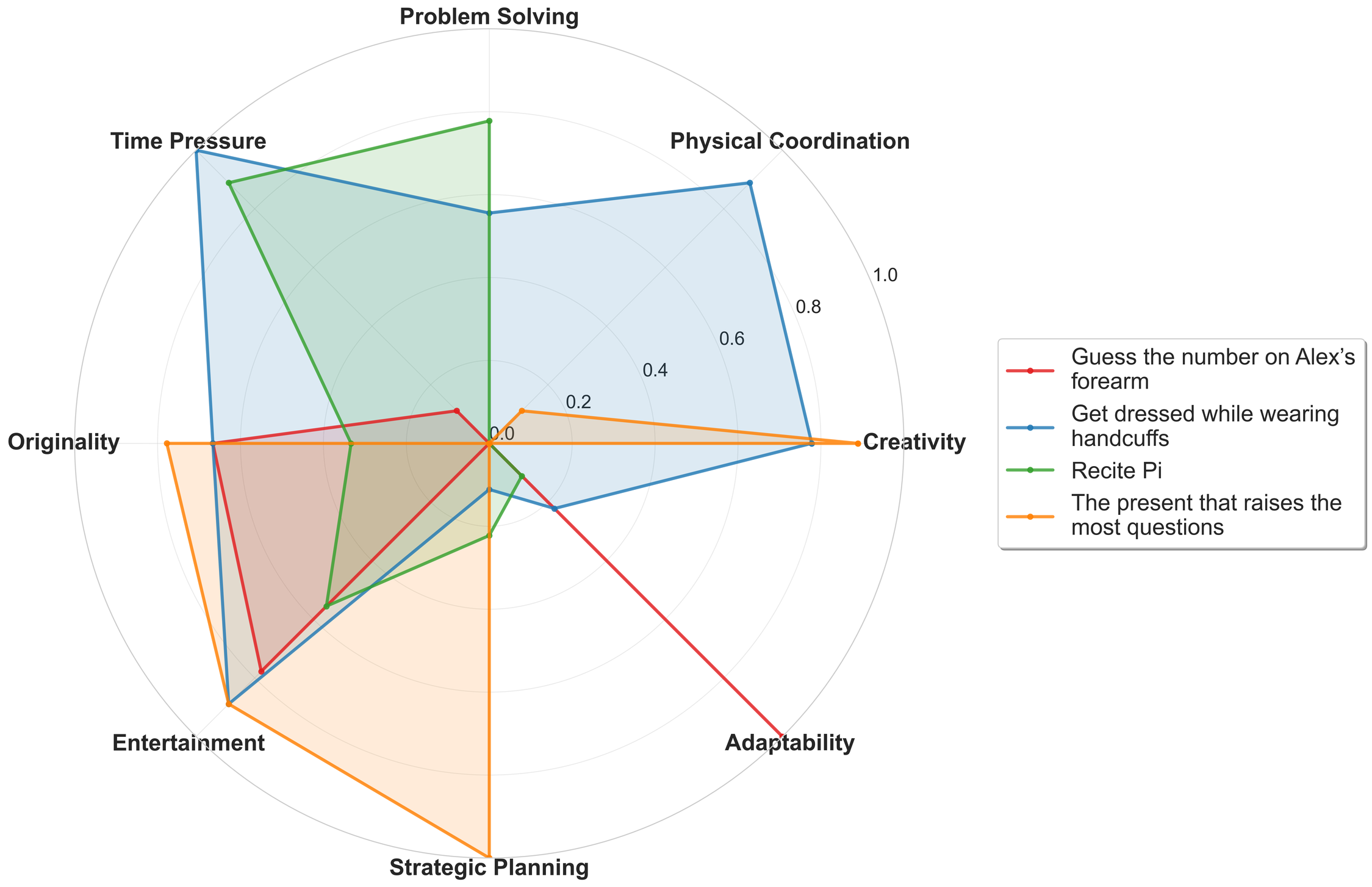}
\caption{{\bf Geographic Origins of Taskmaster Contestants.}
A heatmap showing the birthplaces of all 90 Taskmaster contestants, based on manually collected data and geocoded city-level coordinates. The strongest concentration appears in London, with notable secondary clusters across the UK, Ireland, and a small number of international origins. Base map from Wikishire under Creative Commons Attribution-Share Alike 4.0 International license, original copyright 2014.}
\label{fig:geo_origins}
\end{figure}

To support geographic analyses, birthplaces were geocoded using a public longitude–latitude API. When necessary, city names were disambiguated using administrative region or country context. Of the 90 contestants, 77 (85.6\%) were born in the United Kingdom or Ireland, with the remainder representing international backgrounds (Fig~\ref{fig:geo_origins}). Coordinate data were used in generating regional heatmaps and in computing diversity metrics across series.

\FloatBarrier

\paragraph{Sentiment Annotations}
To quantify emotional tone and comedic texture across episodes, I conducted sentiment analysis using full episode transcripts and GPT-4o, a state-of-the-art large language model. Transcripts were segmented into individual sentences, each of which was annotated with one of seven sentiment categories: awkwardness, humor, joy, sarcasm, anger, self-deprecation, and frustration. Scores were then aggregated at the episode level by averaging across all sentence-level annotations.

The use of automated sentiment analysis is well established in entertainment and media studies. Earlier approaches included lexicon-based tools such as VADER and TextBlob, followed by deep learning models such as BERT \cite{Hutto2014,Devlin2018}. Compared to these methods, GPT-4o offers higher classification accuracy and greater semantic nuance. Benchmark evaluations report a 95.3\% accuracy rate for GPT-4 on sentiment tasks across diverse domains, substantially outperforming classical machine learning models (66–71\%) and lexicon-based approaches (37.2\%) \cite{Fu2024, Debess2024}.

Given this performance, I treated GPT-4o sentiment outputs as reliable first-order measurements of episode-level emotional tone. These values were used descriptively to track trends over time and analytically in correlational studies. This approach provides scalability and consistency across the full 154-episode corpus and aligns with prevailing standards in computational media research.

\subsection*{Task Classification and Format Stability}

To characterize the design of the show’s competitive gameplay, all 917 tasks were categorized along two orthogonal dimensions: \textit{Activity Type} (Creative vs. Physical) and \textit{Judgment Type} (Objective vs. Subjective). This yields a 2×2 matrix of task types, reflecting both the nature of the challenge and the criteria used for evaluation. Labeling was conducted via crowd-sourced annotation and verified through consensus among three independent human raters.

Beyond this core taxonomy, tasks were tagged with additional categorical attributes, including assignment type (solo, team, or split), modality (filmed, live, prize, or homework), and presence or absence of explicit time pressure. This extended metadata framework enabled longitudinal analysis of design choices and stability across all 18 series.

\begin{figure}[!h]
\centering
\includegraphics[width=\linewidth]{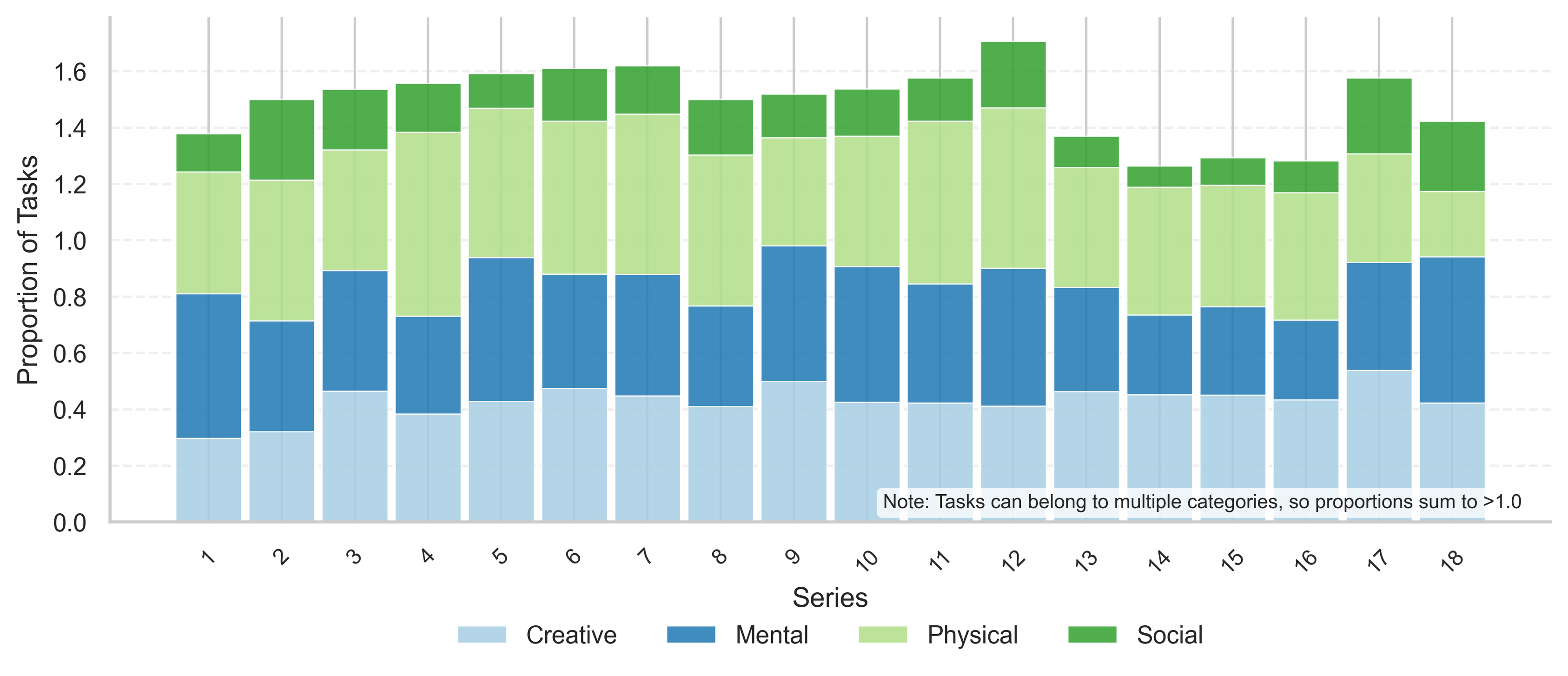}
\caption{{\bf Task Type Distribution Across Series.}
Each of the 917 competitive tasks was labeled by activity type (Creative or Physical) and judgment type (Objective or Subjective). Physical–Objective tasks were most common overall. The proportions of all four task categories remained stable over the show’s 18 series, suggesting a consistent and deliberate design philosophy.}
\label{fig:task_format}
\end{figure}

The quadrant-level distribution of task types was as follows (Fig~\ref{fig:task_format}):
\begin{itemize}
  \item \textbf{Physical–Objective}: 258 tasks (28.1\%). These typically involved concrete, measurable outcomes and physically grounded goals.
  \item \textbf{Creative–Objective}: 182 tasks (19.9\%). These featured novel or artistic outputs assessed using consistent evaluation criteria.
  \item \textbf{Creative–Subjective}: 196 tasks (21.4\%). These relied on judge perception or comedic effect as scoring standards.
  \item \textbf{Physical–Subjective}: 150 tasks (16.4\%). These involved physical activity but were evaluated qualitatively.
\end{itemize}

Despite year-to-year changes in contestants, tone, and production emphasis, the relative proportions of task types remained consistent across series. A temporal trend analysis using Kendall’s $\tau$ found no significant directional shifts in the frequency of any task category ($|\tau| < 0.30$, $p > 0.08$ for all comparisons). This design stability suggests an intentional balance between novelty and familiarity, supporting both comedic variability and viewer expectations.

\FloatBarrier


\section*{Results}

\subsection*{Episode Trajectory Extraction and Pattern Encoding}

To study temporal dynamics in viewer perception, I extracted IMDb ratings for each episode and aligned episodes by ordinal position within each series. Ratings were then normalized within series by subtracting the series mean, yielding a relative rating trajectory for each series. Based on the relative values of the first, middle, and final episodes, I categorized trajectory shapes into one of four archetypes: Rising (123), J-shaped (213), Declining (321), or Flat (no significant difference). Series were assigned to a trajectory type using triplet ordinal comparisons.

\begin{figure}[!h]
\centering
\includegraphics[width=\linewidth]{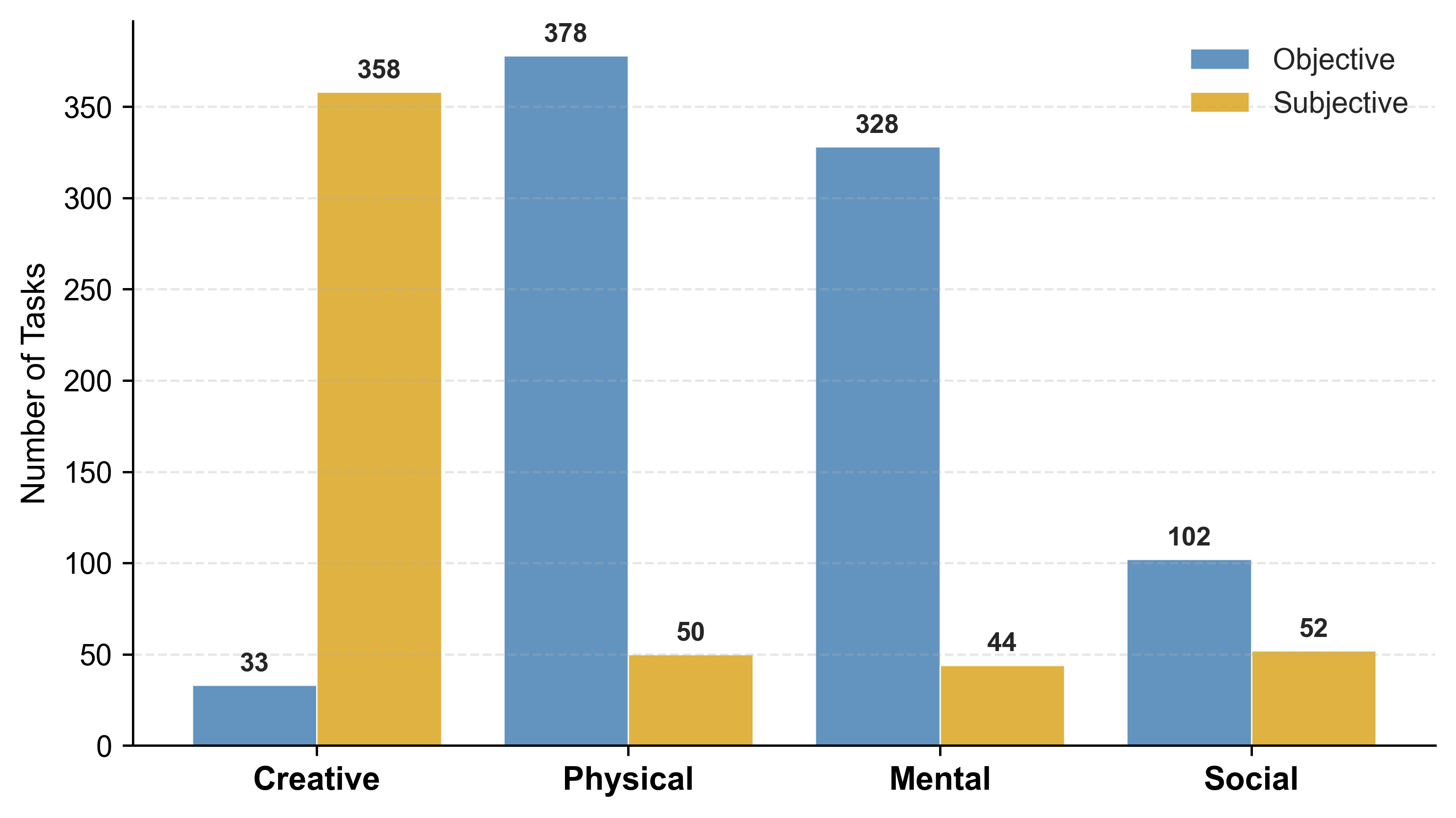}
\caption{{\bf Temporal Patterns of Episode Ratings.}
Episode-level IMDb ratings tend to rise within each series. A majority of series follow either a Rising or J-shaped pattern, with final episodes significantly outperforming first episodes on average ($\Delta = +0.28$, $p < 0.001$). These consistent patterns suggest an emergent narrative across series.}
\label{fig:episode_trajectories}
\end{figure}

Statistical enrichment analysis confirmed the dominance of Rising and J-shaped patterns (Fig~\ref{fig:episode_trajectories}). Sixteen of the eighteen UK series followed one of these two profiles, a highly non-random result supported by both binomial testing ($p < 10^{-6}$) and a chi-square goodness-of-fit test ($\chi^2 = 10.89$, $p = 0.012$). In addition, paired $t$-tests demonstrated that final episodes were consistently rated higher than initial episodes, with an average gain of 0.28 points ($t(17) = 4.92$, $p < 0.001$). This upward trajectory suggests a consistent narrative rhythm in audience reception, likely driven by increased familiarity with the cast, resolution of running jokes, and stronger closers in editing.

These patterns mirror psychological models of expectation and payoff in serialized media, and parallel prior findings in entertainment research where endings tend to anchor final impressions \cite{Fredrickson1993, Vorderer2004, Rozin2004}.

\FloatBarrier

\subsection*{Performance Archetype Assignment via Feature Matching}

To identify recurring patterns in contestant performance, I extracted a 15-dimensional feature vector for each contestant using their cumulative score trajectory and ranking evolution across tasks. Features included early and late average scores, score growth rate, rank variance, rank trajectory volatility, score consistency, comeback factor, and acceleration.

Instead of clustering contestants globally or assuming a fixed archetype count, I defined five idealized archetypes characterized by distinctive performance behaviors:
\begin{enumerate}
  \item \textbf{Steady Performer}: consistent high rankings, low volatility
  \item \textbf{Late Bloomer}: low early scores with improving trend and strong finish
  \item \textbf{Early Star}: strong start but plateau or decline
  \item \textbf{Chaotic Wildcard}: highly variable ranks and scores
  \item \textbf{Consistent Middle}: average performance, little fluctuation
\end{enumerate}

Each archetype was formalized as a weighted scoring function over the standardized features. This approach reflects principles in cluster interpretation where silhouette-based methods are used to evaluate the coherence and distinctiveness of clusters~\cite{Rousseeuw1987}. For example, Late Bloomers were optimized for high late score ratio, strong score growth, and a large comeback factor. Within each series, I greedily assigned one contestant to each archetype by selecting the contestant with the highest score under each archetype’s definition, ensuring a perfect 1:1 mapping between contestants and archetypes per series.

\begin{figure}[!h]
\centering
\includegraphics[width=\linewidth]{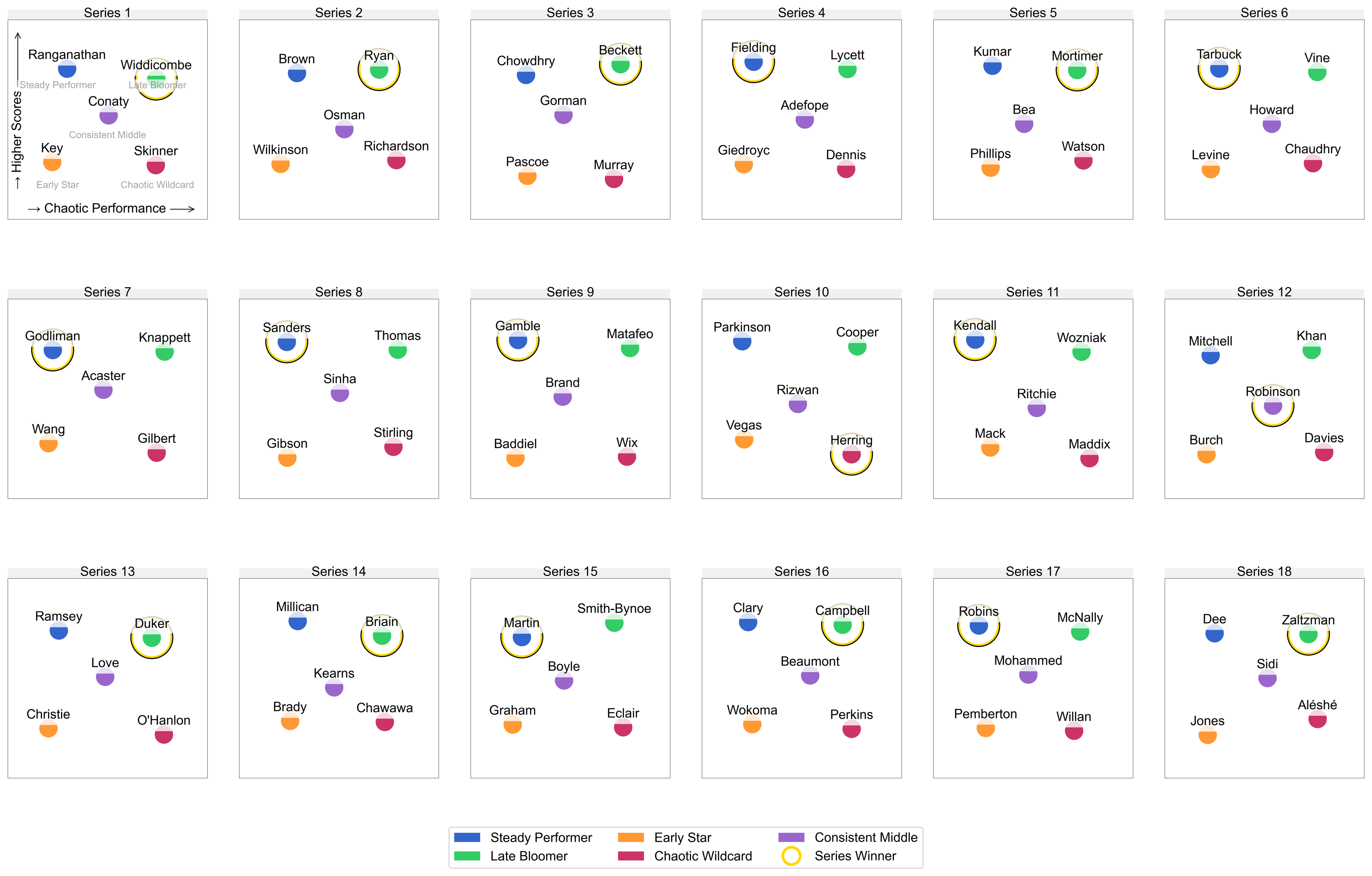}
\caption{{\bf Performance Archetypes Across Series.}
Contestant scoring trajectories were clustered into five recurring archetypes based on their task-by-task performance patterns. Each series features one representative of each type. The most common winning profiles are the Steady Performer (top-left), characterized by consistent scoring, and the Late Bloomer (top-right), who improves substantially over time. Other archetypes include the Early Star (bottom-left), who starts strong but declines in later episodes; the Chaotic Wildcard (bottom-right), marked by erratic and unpredictable performance; and the Consistent Middle (center), who maintains a moderate rank throughout without major highs or lows.}
\label{fig:performance_archetypes}
\end{figure}

This optimization ensures that each series contains one archetypal representative of each performance mode, allowing us to examine how narrative roles repeat across seasons (Fig~\ref{fig:performance_archetypes}). Final placement outcomes aligned strongly with archetype identity: 66\% of series winners were either Late Bloomers or Steady Performers, while Consistent Middles finished last in 44\% of series.

\FloatBarrier

\subsection*{Scoring Geometry and Pattern Usage}

In each competitive task, five contestants receive integer scores between 0 and 5, typically determined by the show's judge based on performance or creativity. This creates a five-element vector of scores per task. Given the discrete and bounded nature of this scoring system, the total number of mathematically valid permutations is finite. After applying \textit{Taskmaster}'s rules (e.g., no duplicate scores above five, adherence to task-specific fairness), I enumerated all 252 unique five-contestant score distributions.

To characterize these scoring patterns and measure their diversity, I explored various ways of comparing score vectors. Distance metrics over full distributions (e.g., Euclidean or Wasserstein distances) proved difficult to interpret and did not capture meaningful features. In contrast, basic statistical moments—specifically, the mean ($\mu$), variance ($\sigma^2$), and skewness of each vector—offered a compact and interpretable embedding of each pattern into a three-dimensional space. These metrics captured overall generosity, score spread, and asymmetry in scoring without requiring arbitrary distance choices. The approach of summarizing complex distributions using a small number of interpretable statistics follows foundational work on dimension reduction and mixture modeling in statistics~\cite{Pearson1901,Hotelling1933,McLachlan2000}.

\begin{figure}[!h]
\centering
\includegraphics[width=\linewidth]{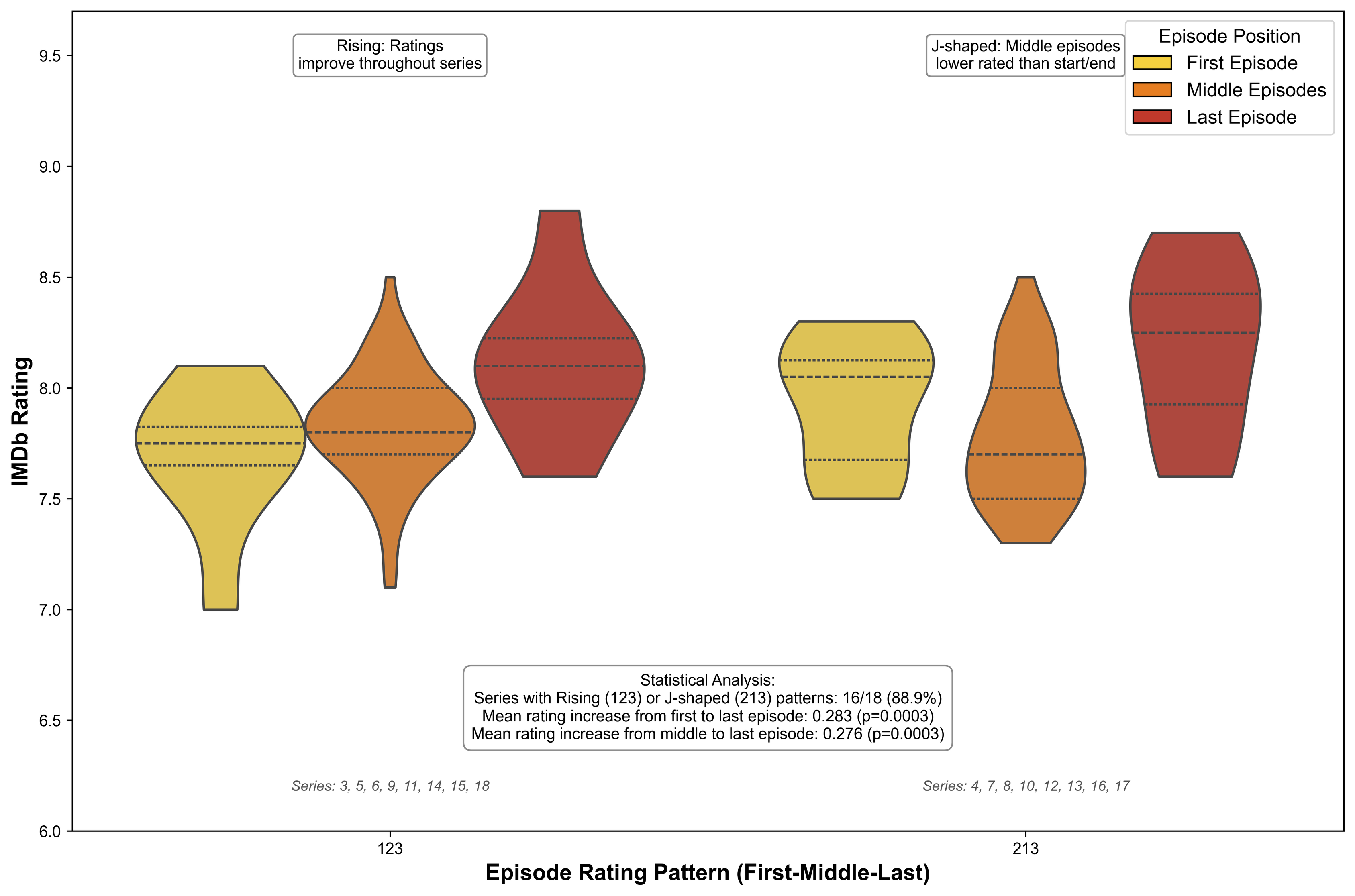}
\caption{{\bf Geometry of Task Scoring Patterns.}
Each point represents a unique five-contestant scoring pattern, positioned by its mean, variance, and skewness. Black points indicate patterns used in the show; gray points are unused configurations. Used patterns cluster in a mid-range region, suggesting a preference for scoring dynamics that promote moderate competitiveness and differentiation.}
\label{fig:scoring_geometry}
\end{figure}

We then mapped each task in the dataset to its corresponding pattern and projected all 252 possible patterns into the resulting $\mu$–$\sigma^2$–skewness space. Out of the 252 mathematically possible configurations, only 98 (38.9\%) were ever used in the show (Fig~\ref{fig:scoring_geometry}). These used patterns tended to cluster around a central band with moderate generosity (mean $\mu \approx 3$) and moderate spread (variance $\approx 2$). Unused patterns included highly uniform distributions (e.g., all contestants tied with the same score) or highly skewed, extreme cases. This suggests that producers and editors, whether implicitly or intentionally, favor scoring outcomes that support narrative variety and momentum, while avoiding both excessive randomness and complete uniformity.

\FloatBarrier

\subsection*{Sentiment Annotation and Temporal Analysis}

We performed episode-level sentiment extraction across seven emotion dimensions: awkwardness, humor, anger, joy/excitement, sarcasm, frustration, and self-deprecation. Sentiment scores were computed by averaging per-sentence annotations within each transcript-parsed episode.

\begin{figure}[!h]
\centering
\includegraphics[width=\linewidth]{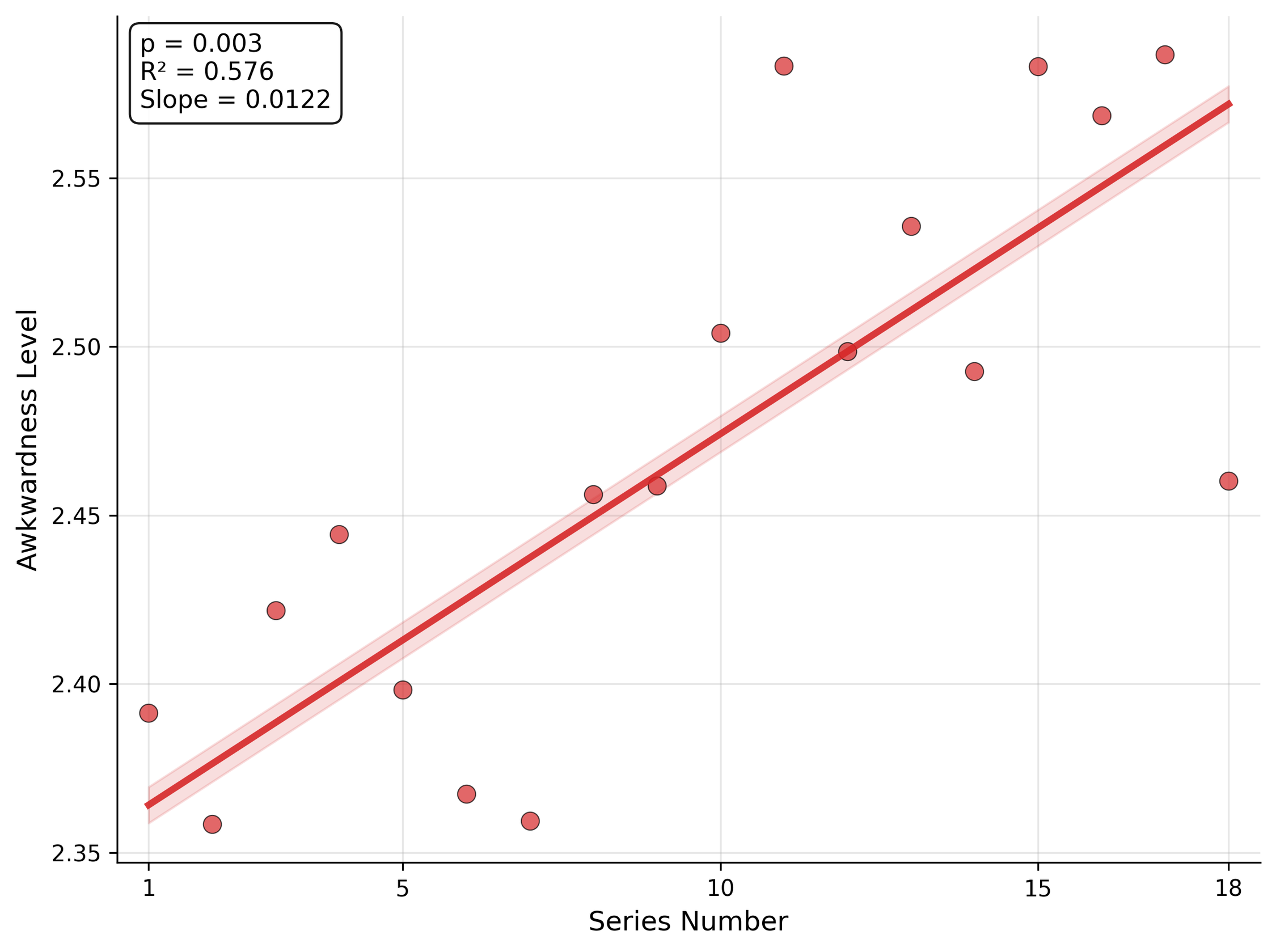}
\caption{{\bf Sentiment Trends Across 18 Series.}
Among seven analyzed sentiment categories, only awkwardness shows a significant increase over time. The upward trend suggests a shift in comedic tone toward more discomfort-based humor, consistent with broader trends in contemporary media.}
\label{fig:sentiment_trends}
\end{figure}

Among these dimensions, only awkwardness showed a statistically significant longitudinal trend across series (Fig~\ref{fig:sentiment_trends}). A linear regression revealed a strong positive effect ($\beta = 0.0122$, $p = 0.0004$), which remained significant after False Discovery Rate correction~\cite{Benjamini1995} (adjusted $p = 0.0027$). The standardized effect size was large (Cohen’s $d = 2.71$), suggesting that more recent series have leaned increasingly into “cringe” or discomfort-based humor. No significant trends were observed for the remaining sentiment categories, indicating overall emotional tone stability across the show’s lifespan.

\FloatBarrier

\subsection*{Predictive Modeling and Feature Attribution}

To assess which features best predict audience reception, I trained Random Forest regressors to model IMDb episode ratings using 45 features drawn from three domains: contestant demographics, task structure, and sentiment tone. Feature construction included variables such as average age, gender composition, professional background, task proportions, and episode-level sentiment scores. All models were evaluated using 5-fold cross-validation, with splits performed at the series level to reduce overfitting to stylistic or tonal patterns within any given season \cite{Arlot2010}.

\begin{figure}[!h]
\centering
\includegraphics[width=\linewidth]{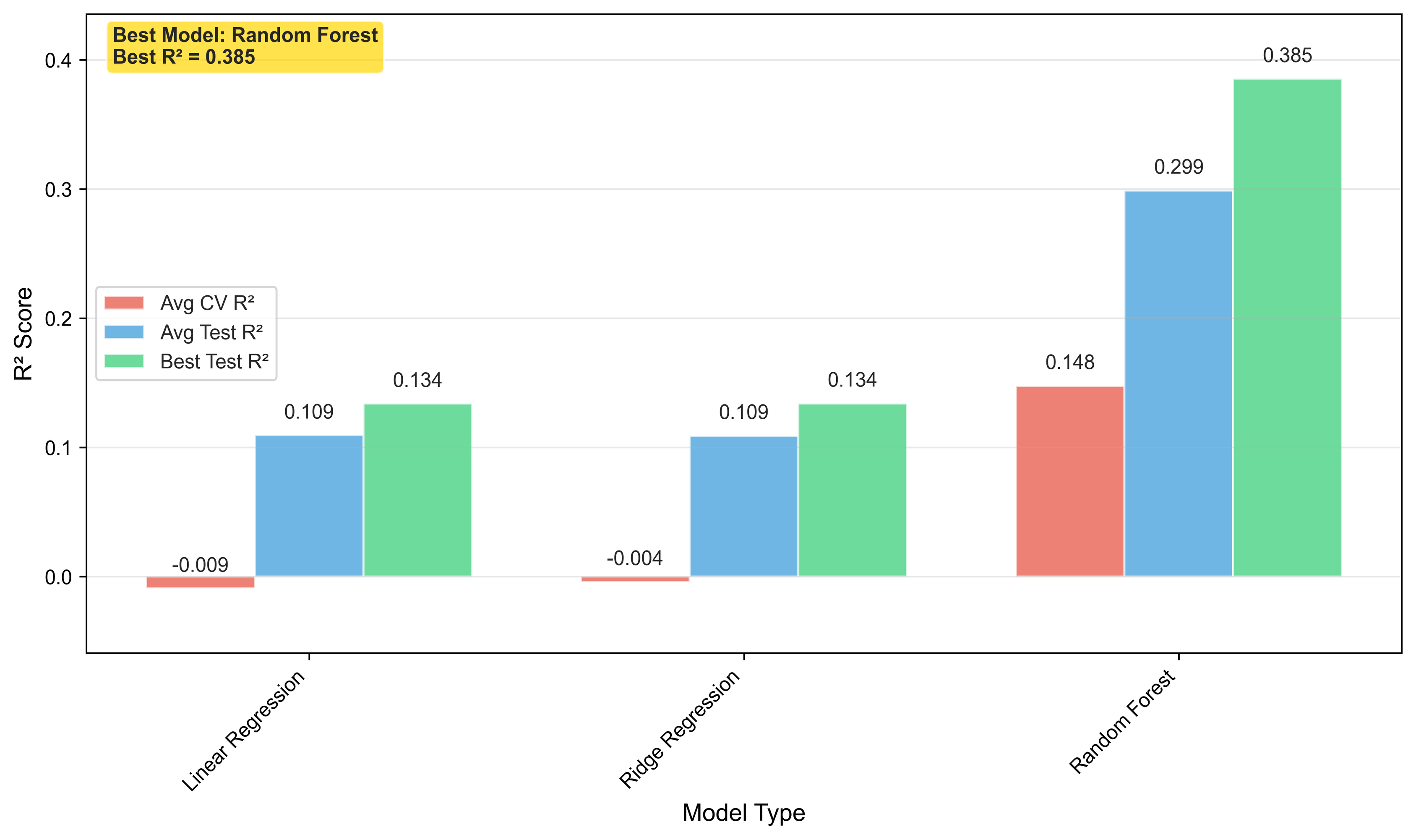}
\caption{{\bf Predicting Episode Ratings Using Machine Learning.}
A Random Forest model predicts IMDb episode ratings using 45 features derived from contestant demographics, task attributes, and emotional tone. Cross-validation was performed by holding out entire series to prevent overfitting to seasonal tone or cast-specific patterns. The strongest predictors were contestant age, prior experience, and mean awkwardness. All major features showed positive associations with ratings, except awkwardness, which was negatively correlated. These three features together accounted for over 88\% of total model importance.}
\label{fig:ml_prediction}
\end{figure}

The best-performing model was a Random Forest with 500 estimators and a maximum tree depth of 5 \cite{Breiman2001} (Fig~\ref{fig:ml_prediction}). It achieved a cross-validated $R^2 = 0.385$, indicating moderate predictive performance. By holding out entire series during validation, I ensured that the model's predictions generalized across seasons, rather than capturing idiosyncrasies of individual episodes or casts.

\begin{figure}[!h]
\centering
\includegraphics[width=\linewidth]{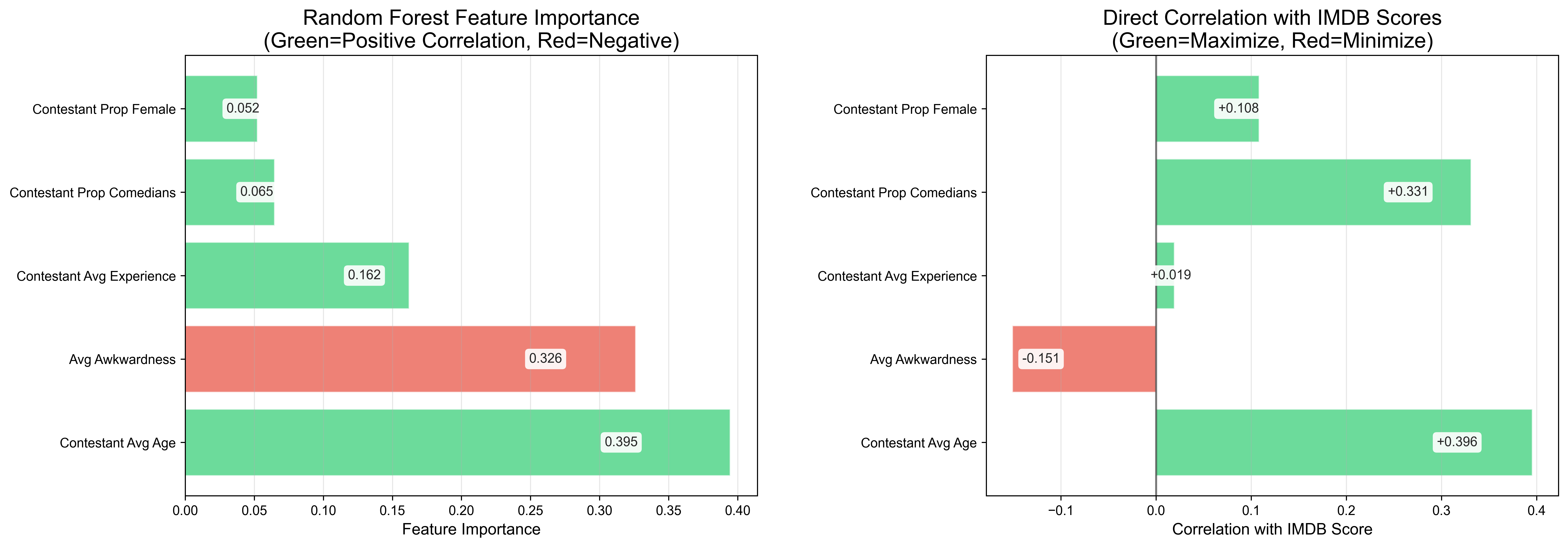}
\caption{{\bf Feature Importance Breakdown.}
Relative feature importances from the Random Forest model, sorted from most to least predictive. Contestant-level features, particularly age, awkwardness, and experience, dominate. Task-related and scoring dynamics variables have minimal influence on predicted ratings.}
\label{fig:feature_importance}
\end{figure}

Feature attribution analysis revealed that contestant-related variables accounted for over 88\% of total model importance (Fig~\ref{fig:feature_importance}). All major predictors were positively associated with IMDb scores, with the exception of awkwardness, which was negatively correlated:
\begin{itemize}
  \item \textbf{Contestant average age} (39.5\%). Older casts were associated with higher ratings.
  \item \textbf{Mean awkwardness} (32.6\%). Increased awkwardness was negatively associated with audience reception.
  \item \textbf{Performer experience (years)} (16.2\%). More experienced performers tended to boost ratings.
  \item \textbf{Comedian proportion} (6.5\%). Lineups featuring more professional comedians correlated positively with ratings.
\end{itemize}

These results suggest that the strongest predictors of audience response are performer attributes rather than episode flow. The data point to an ideal cast configuration that includes experienced comedians, some over 40 years old with minimal awkwardness. A complementary series-level correlation analysis ($n = 18$) confirmed that actor-heavy lineups were negatively associated with mean ratings ($r = -0.547$), while greater proportions of comedians and solo-task formats correlated positively ($r > 0.33$).

Together, these findings support the conclusion that viewer satisfaction in hybrid entertainment formats is shaped more by the polish, composure, and tone of performer interactions than by competitive mechanics or scoring complexity.

\FloatBarrier

\FloatBarrier


\section*{Discussion}

This study offers a quantitative examination of audience engagement in a hybrid entertainment format that blends comedy with structured competition. Using 154 episodes of \textit{Taskmaster UK}, I analyzed over 900 tasks and more than 32,000 IMDb ratings to investigate how scoring dynamics, contestant attributes, and emotional tone relate to viewer perception. Our results highlight several key insights about what drives success in this format.

First, our analysis of episode rating trajectories revealed strong evidence for a consistent narrative rhythm across series. All seasons but two followed either a Rising or J-shaped pattern, with episode ratings increasing toward the finale. This suggests that viewers respond positively to gradual character development, familiarity with the cast, and resolution of narrative arcs. These findings echo prior research in entertainment psychology showing that endings exert a disproportionate influence on retrospective evaluations \cite{Fredrickson1993, Rozin2004}.

Second, task design showed a high degree of consistency. The distribution of task types across the Creative/Physical and Objective/Subjective dimensions remained stable across all 18 series. Despite year-to-year changes in tone and contestants, the producers maintained a consistent mix of task formats. This deliberate balance likely supports the show’s longevity by anchoring comedic improvisation within a familiar structure.

Third, scoring analysis revealed that only 39\% of mathematically possible five-player score distributions were ever used. The show's scoring patterns cluster in a moderate range of mean and variance, avoiding both extreme randomness and uniformity. This suggests a preference for scoring configurations that support narrative differentiation without undermining fairness or coherence.

Fourth, sentiment analysis showed a significant increase in awkwardness over time, while other emotional tones remained stable. This supports the interpretation that the show's comedic strategy has shifted toward more discomfort-based humor in recent years, aligning with broader trends in British and online comedy.

Finally, predictive modeling results reinforced the centrality of contestant characteristics in driving audience response, with older, more experienced performers associated with higher ratings while awkwardness negatively correlated with reception.

Despite these insights, several limitations must be acknowledged. Our sentiment annotations, while validated against benchmarks, were derived from transcript-based GPT-4o analyses and may not fully capture multimodal cues such as delivery, timing, or audience laughter. In addition, our scoring geometry approach, while interpretable, does not account for semantic content or judge commentary. Finally, IMDb ratings reflect a non-random and self-selecting sample of viewers, and may not generalize to broader populations.

\section*{Conclusions}

This analysis of \textit{Taskmaster UK} demonstrates that performer characteristics—age, experience, and emotional tone—are the primary determinants of audience engagement in hybrid entertainment formats, while traditional competitive metrics play only a limited role in predicting viewer reception.

Across multiple analytical levels, I observed consistent patterns. Episode ratings tend to rise across each series, suggesting a robust narrative shaped by character familiarity and payoff timing. Scoring patterns remain within a narrow band of used configurations, avoiding both randomness and repetition. Sentiment analysis reveals an increasing reliance on awkwardness over time, pointing to a shift in comedic tone. Most notably, predictive modeling shows that experienced, older comedians consistently correlate with higher ratings, while emotional discomfort detracts from audience perception.

These results suggest that audience engagement in hybrid formats is shaped less by suspense or novelty and more by casting, emotional tone, and the on-screen personas of the contestants. While competition provides the framework, it is the performers’ delivery, polish, and chemistry that determine reception.

\subsection*{Implications for Entertainment Industry Practice}

Our findings offer several actionable insights for television producers, format developers, and entertainment industry executives working with hybrid formats that combine competitive and comedic elements.

\paragraph{Casting and Performer Selection}
Producers should prioritize experienced performers over 40 years old when optimizing for audience engagement. The entertainment industry's emphasis on youth demographics may be misaligned with what drives viewer satisfaction in hybrid formats, where seasoned comedians consistently correlate with higher ratings.

\paragraph{Emotional Tone and Content Management}
While contemporary comedy increasingly embraces "cringe" humor~\cite{HyeKnudsen2018}, excessive emotional discomfort actively detracts from viewer enjoyment. Producers should implement content review processes that monitor and limit awkward moments during post-production editing, challenging the assumption that controversial content automatically generates higher engagement.

\paragraph{Competition Design and Narrative Structure}
Competitive mechanics contribute minimally to audience satisfaction. Producers should focus resources on performer chemistry, comedic writing, and character development rather than elaborate scoring systems or manufactured suspense. Competition should serve as scaffolding for personality-driven content rather than the primary source of engagement.

\paragraph{Series Pacing and Episode Trajectory}
Design seasons to build toward stronger finales, allowing time for character familiarity and comedic chemistry to develop. This contradicts the common practice of front-loading seasons with high-concept episodes, suggesting that gradual escalation and payoff timing are more effective for sustained engagement.

\paragraph{Format Adaptation and International Considerations}
Ensure casting includes representatives of performance archetypes: Steady Performer, Late Bloomer, Early Star, Chaotic Wildcard, and Consistent Middle. This archetypal diversity provides narrative balance and viewer identification opportunities that transcend cultural boundaries.

Future work could extend this approach to a broader comparative framework. By extracting a compact set of descriptors for each show—such as the degree of scoring centrality, comedic improvisation, emotional volatility, or consistency—it may be possible to embed entire formats into a dynamic feature space. Mapping these embeddings could help reveal genre boundaries, latent archetypes, or evolutionary paths in television entertainment, offering both theoretical insight and practical guidance for producers and distributors.

\section*{Acknowledgments and Declarations}

I thank the contributors and maintainers of \href{https://taskmaster.info}{Taskmaster.info}, whose structured episode and task archive made this analysis easier. I am also grateful to Alex Horne and Greg Davies for creating and sustaining a show that is not only culturally rich but uniquely amenable to data-driven analysis.

The author declares no funding sources and no competing interests.


%
%
%

\section*{Supporting information}

\paragraph*{S1 Fig.}
{\bf Sentiment Variability Across Episodes.} 
Box plots showing the distribution of each sentiment category across the full 154-episode corpus. While mean levels remain stable, awkwardness exhibits a broader distribution in later series.

\paragraph*{S2 Fig.}
{\bf Feature–Rating Correlation Histogram.} 
Distribution of Pearson correlation coefficients between 45 episode-level features and IMDb ratings. Sentiment and contestant attributes dominate the upper tail, while task features are near zero.

\paragraph*{S3 Fig.}
{\bf Series-Level Deep Dive Visualizations.}
One-page summaries of all 18 Taskmaster UK series. Each page includes two panels: cumulative contestant score progression (top) and per-task rank evolution (bottom). These visualizations support archetype consistency and highlight turning points in series narratives.


\begin{thebibliography}{22}

\bibitem{Gray2009}
Gray J, Jones JP, Thompson E.
\newblock Satire TV: Politics and comedy in the post-network era.
\newblock New York: NYU Press; 2009.

\bibitem{Katz1973}
Katz E, Blumler JG, Gurevitch M.
\newblock Uses and gratifications research.
\newblock Public Opin Q. 1973;37(4):509--523.
\newblock doi:10.1086/268109

\bibitem{Hill2017}
Hill A.
\newblock Reality TV engagement: Producer and audience relations for reality talent shows.
\newblock Media Ind J. 2017;4(1):123--145.
\newblock doi:10.3998/mij.15031809.0004.106

\bibitem{Enli2012}
Enli GS.
\newblock From parasocial interaction to social TV: Analysing the host–audience relationship in multi-platform productions.
\newblock North Lights. 2012;10(1):123--137.
\newblock doi:10.1386/nl.10.1.123\_1

\bibitem{Rubin2009}
Rubin AM.
\newblock Uses-and-gratifications perspective on media effects.
\newblock In: Bryant J, Oliver MB, editors. Media effects: Advances in theory and research. 3rd ed. New York: Routledge; 2009. p. 165--184.

\bibitem{Nabi2003}
Nabi RL, Biely EN, Morgan SJ, Stitt CR.
\newblock Reality-based television programming and the psychology of its appeal.
\newblock Media Psychol. 2003;5(4):303--330.
\newblock doi:10.1207/S1532785XMEP0504\_01

\bibitem{GarciaBéjar2021}
Benavides Almarza C, García-Béjar L.
\newblock ¿Por qué ven Netflix quienes ven Netflix?: experiencias de engagement de jóvenes mexicanos frente a quien revolucionó el consumo audiovisual.
\newblock Rev Comun. 2021;20(1):29--47.
\newblock doi:10.26441/RC20.1-2021-A2

\bibitem{Vassallo2016}
Vassallo de Lopes MI.
\newblock (Re)Invention of TV Fiction: Genres and Formats.
\newblock OBITEL 2016. Porto Alegre: Sulina; 2016.

\bibitem{Gil2012}
Gil E.
\newblock Telefantasy's Conflicting Verisimilitudes: Composite Genre.
\newblock Sci Fiction Film Telev. 2012;5(2):175--193.
\newblock doi:10.3828/sfftv.2012.10

\bibitem{OBITEL2018}
OBITEL.
\newblock Spain: Innovation and hybridization of genres and formats.
\newblock OBITEL 2018: Audiovisual distribution on the internet; 2018.

\bibitem{Larkey2016}
Larkey E, Schwarzenegger C, Žáková E.
\newblock Measuring Transnationalism: Comparing TV Formats using Digital Tools.
\newblock VIEW J Eur Telev Hist Cult. 2016;5(9):67--83.
\newblock doi:10.18146/2213-0969.2016.jethc103

\bibitem{Straubhaar2007}
Straubhaar J.
\newblock World Television: From Global to Local.
\newblock 2nd ed. Thousand Oaks: Sage Publications; 2007.

\bibitem{Trepte2008}
Trepte S.
\newblock Cultural proximity in TV entertainment: An eight-country study on the relationship of nationality and the evaluation of U.S. prime-time fiction.
\newblock Commun Res. 2008;35(4):506--524.
\newblock doi:10.1177/0093650208316317

\bibitem{YooEtAl2014}
Yoo JW, Mackenzie NG, Jones VM.
\newblock Effects of television viewing, cultural proximity, and ethnocentrism on country image.
\newblock Soc Behav Personal. 2014;42(1):105--116.
\newblock doi:10.2224/sbp.2014.42.1.105

\bibitem{McLachlan2000}
McLachlan G, Peel D.
\newblock Finite mixture models.
\newblock New York: Wiley; 2000.

\bibitem{Rottenberg1956}
Rottenberg S.
\newblock The baseball players’ labor market.
\newblock J Polit Econ. 1956;64(3):242--258.

\bibitem{Fort2003}
Fort R, Maxcy J.
\newblock Competitive balance in sports leagues: an introduction.
\newblock J Sports Econ. 2003;4(2):154--160.

\bibitem{Hall2009}
Hall A.
\newblock Viewer perceptions of reality programs.
\newblock Media Psychol. 2009;12(2):181--195.

\bibitem{Vorderer2004}
Vorderer P, Klimmt C, Ritterfeld U.
\newblock Enjoyment: at the heart of media entertainment.
\newblock Commun Theory. 2004;14(4):388--408.

\bibitem{Gelman2007}
Gelman A, Hill J.
\newblock Data analysis using regression and multilevel/hierarchical models.
\newblock Cambridge: Cambridge University Press; 2007.

\bibitem{imdb}
Internet Movie Database (IMDb): Episode Rating Pages;
\newblock 2024. Available from: \url{https://www.imdb.com}.

\bibitem{taskmasterinfo}
Taskmaster.info: Episode and Task Archive;
\newblock 2024. Available from: \url{https://taskmaster.info}.

\bibitem{Hutto2014}
Hutto CJ, Gilbert E.
\newblock VADER: A parsimonious rule-based model for sentiment analysis of social media text.
\newblock In: Proceedings of the Eighth International Conference on Weblogs and Social Media (ICWSM); 2014.

\bibitem{HyeKnudsen2018}
Hye-Knudsen M.
\newblock Painfully Funny: Cringe Comedy, Benign Masochism, and Not-So-Benign Violations.
\newblock Leviathan: Interdisciplinary Journal in English. 2018;(2):13--31.
\newblock doi:10.7146/lev.v0i2.104693

\bibitem{Devlin2018}
Devlin J, Chang MW, Lee K, Toutanova K.
\newblock BERT: Pre-training of deep bidirectional transformers for language understanding.
\newblock arXiv preprint arXiv:1810.04805. 2018.

\bibitem{Fu2024}
Fu Z, Hsu YC, Chan CS, Lau CM, Liu J, Yip PSF.
\newblock Efficacy of ChatGPT in Cantonese Sentiment Analysis: Comparative Study.
\newblock J Med Internet Res. 2024;26:e51069.
\newblock doi:10.2196/51069

\bibitem{Debess2024}
Debess IN, Simonsen A, Einarsson H.
\newblock Good or Bad News? Exploring GPT-4 for Sentiment Analysis for Faroese on a Public News Corpora.
\newblock In: Proceedings of the 2024 Joint International Conference on Computational Linguistics, Language Resources and Evaluation (LREC-COLING 2024); 2024.
\newblock Available from: \url{https://aclanthology.org/2024.lrec-main.690/}.

\bibitem{Fredrickson1993}
Fredrickson BL, Kahneman D.
\newblock Duration neglect in retrospective evaluations of affective episodes.
\newblock J Pers Soc Psychol. 1993;65(1):45--55.
\newblock doi:10.1037/0022-3514.65.1.45


\bibitem{Rozin2004}
Rozin P, Royzman EB.
\newblock Temporal order and retrospective evaluations of emotional episodes.
\newblock J Behav Decis Mak. 2004;17(1):1--12.

\bibitem{Arlot2010}
Arlot S, Celisse A.
\newblock A survey of cross-validation procedures for model selection.
\newblock Stat Surv. 2010;4:40--79.

\bibitem{Breiman2001}
Breiman L.
\newblock Random forests.
\newblock Mach Learn. 2001;45(1):5--32.

\bibitem{Benjamini1995}
Benjamini Y, Hochberg Y.
\newblock Controlling the false discovery rate: a practical and powerful approach to multiple testing.
\newblock J R Stat Soc Ser B. 1995;57(1):289--300.

\bibitem{Rousseeuw1987}
Rousseeuw PJ.
\newblock Silhouettes: a graphical aid to the interpretation and validation of cluster analysis.
\newblock J Comput Appl Math. 1987;20:53--65.

\bibitem{Pearson1901}
Pearson K.
\newblock On lines and planes of closest fit to systems of points in space.
\newblock Philos Mag. 1901;2(11):559--572.

\bibitem{Hotelling1933}
Hotelling H.
\newblock Analysis of a complex of statistical variables into principal components.
\newblock J Educ Psychol. 1933;24(6):417--441.

\bibitem{Hastie2009}
Hastie T, Tibshirani R, Friedman J.
\newblock The elements of statistical learning. 2nd ed.
\newblock New York: Springer; 2009.

\bibitem{James2013}
James G, Witten D, Hastie T, Tibshirani R.
\newblock An introduction to statistical learning.
\newblock New York: Springer; 2013.

\end{thebibliography}
\end{document}